\documentclass[10pt]{iopart}
\usepackage{epsfig}
\usepackage{supertabular}

\begin{document}
\title{Right on time: Measuring Kuramoto model coupling from a survey of wrist-watches}
\author{Reginald D. Smith}

\address{Bouchet-Franklin Research Institute PO Box 10051 Rochester NY 14610}
\ead{rsmith@bouchet-franklin.org}
\date{March 25, 2010}

\begin{abstract}
Using a survey of wrist-watch synchronization from a randomly selected group of independent volunteers, we model the system as a Kuramoto-type coupled oscillator network. Based on the phase data both the order parameter and likely size of the coupling is derived and the possibilities for similar research to deduce topology from dynamics are discussed.


PACC: 6460C, 0545

\end{abstract}

\maketitle
\section{Introduction}

One of the most pressing problems in complex systems and science of complex networks is the relationship between network topology and the dynamics which operate over the network topology over time. For one, the growth and development of the topology and dynamics, while undoubtedly linked, usually occur on vastly separated timescales with fluctuations of the internal dynamics changing rapidly over short timescales, though perhaps with long-term trends, and with the topology often growing and developing over timescales much longer than those influenced by the key dynamic drivers. The complete answer to this question is still unresolved.

One of the most promising avenues for investigating the relationship between topology and dynamics is the synthesis of complex network topologies with network oscillator synchronization models such as those based on the well-studied Kuramoto coupled oscillator mean-field models. The model's basic development is well chronicled by Strogatz in \cite{strogatzhistory} from Winfree's first mathematical exposition to the work by Kuramoto on the dominant model of collective synchronization. In a series of papers, \cite{kuramoto1,kuramoto2,kuramoto3,kuramoto4} Kuramoto developed a mean-field solution for a system of all-to-all coupled oscillators coupled according to the relation

\begin{equation}
\dot{\theta_i} = \omega_i +\frac{K}{N}\sum^N_{j=1} \sin(\theta_j - \theta_i), i = 1\ldots N
\label{basic}
\end{equation}

coupled through a global coupling strength $K$. What was realized that by increasing $K$ past a critical threshold, $K_c$, the oscillators began to exhibit collective synchronization and oscillate in phase. The phase synchronization of the oscillators is measured by the order parameter, $r$, 

\begin{equation}
re^{i\psi}=\frac{1}{N}\sum^N_{j=1}e^{i\theta_j}
\label{r}
\end{equation}

where $\theta_i$ is the phase of $i$th oscillator and $\psi$ is the average phase among all the oscillators. In the model, the synchronization is represented by $r  = 0$ for $K<K_c$ and $r>0$ for $K\geq K_c$. 

The development and analysis of the Kuramoto model usually follows an exposition of the synchronization dynamics of the oscillators using theoretical argument or simulations. A particular focus is the value of $K_c$ or a plot of $r$ vs. $K$, under various conditions such as scale-free network topologies \cite{scalefree} or modified dynamics. For a review see \cite{netsyncreview}. 

In this paper, the reverse approach will be investigated, in particular, using real-world data from a hypothesized oscillator network to back out characteristics of the network coupling, and hopefully, structure. 

\section{System description \& experimental setup}

In modern society, we take it for granted that everyone knows how to be on time. With the availability of inexpensive, portable, and relatively accurate timepieces such as watches and increasingly cell phones and PDAs, the fact that everyone's watch is ``synchronized'' to a certain extent is of seemingly trivial interest. However, historically it has not always been so and even though clocks have existed for centuries, modern standard times and time zones only emerged with the advances of long-range telecommunications and transportation in the 19th century. It can be readily acknowledged that there are ``references'' that time can be based on. Accurate time measurements can originate from organizations such as NIST and its broadcasts on 5000 kHz and other related frequencies. However, this is rarely directly accessed by the average person. In addition, the rise of cell phones, which often have their time beamed from the cell-tower's synchronized clock, also helps provide accuracy. However, despite these it is the author's contention that much of the synchronization of timepieces to such close precision is due to a coordination among multiple independent actors who realize that a valuable social consensus ``being on time'' is important and is largely a relative measurement. 

In order to see if this coordination could be measured as a real effect and related to synchronized dynamics, the author performed a survey experiment to measure the synchronization amongst watches for a group of random and independent people. On September 19, 2009 the author set up a survey booth for five hours at the Rochester Public Market in Rochester, NY. The Rochester Public Market is a large and popular open air market selling produce and crafts that is held weekly on Saturdays in Rochester and allowed an easily accessible and random group of strangers. From the survey booth, the author solicited volunteers for a quick experiment. With a laptop and web camera setup, the author quickly explained the goals of the experiment and invited the volunteer to allow his or her watch to have its face captured by the web camera. As mentioned earlier, cell phones were not deemed acceptable since they are automatically synchronized.

The web camera created a JPEG image file which had an exact timestamp recording the time of the picture capture. Once the survey had ended, the time on the face of the watch and its deviation, positive or negative, from the hour and minute on the file timestep was recorded with a precision of  $\pm$ 1 minute. This deviation is the phase difference between the watch and the clock and the collected phase differences amongst all the watches is used to investigate the synchronization of the network. The watches are all assumed to be oscillators with an identical frequency $\omega = \frac{2\pi}{60}$.

It was determined that using the most general model, the Kuramoto model, the mean coupling of the hypothesized oscillator network could be derived to give a general idea of the strength of coupling in the watch synchronization network.

\section{Calculations and results}

\begin{figure}[t]
\includegraphics[height=3in, width=3in]{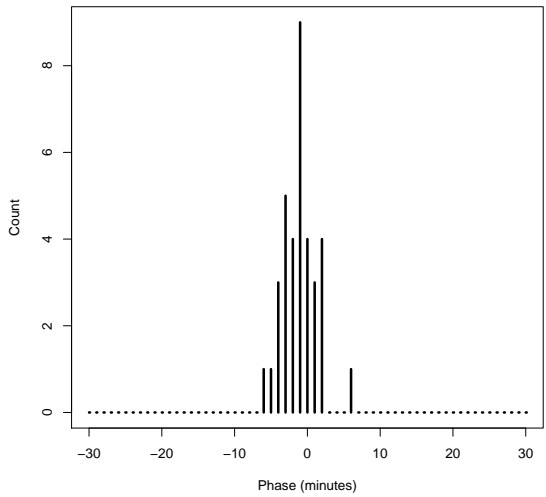}
\caption{The count (not proportional) phase density plot of the measurements of watch time phase difference from the computer timestamp as measured in the experiment.}
\label{timedistribution}
\end{figure}
\begin{figure}[t]
\includegraphics[height=3in, width=3in]{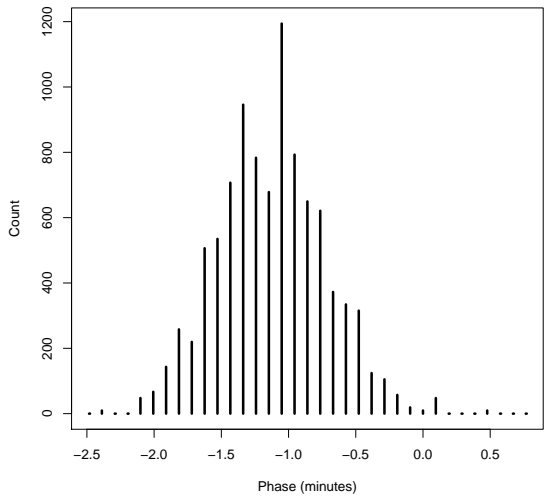}
\caption{The distribution of the mean phase from a 1000 sample bootstrap.}
\label{timedistribution}
\end{figure}

\begin{figure}[t]
\includegraphics[height=3in, width=3in]{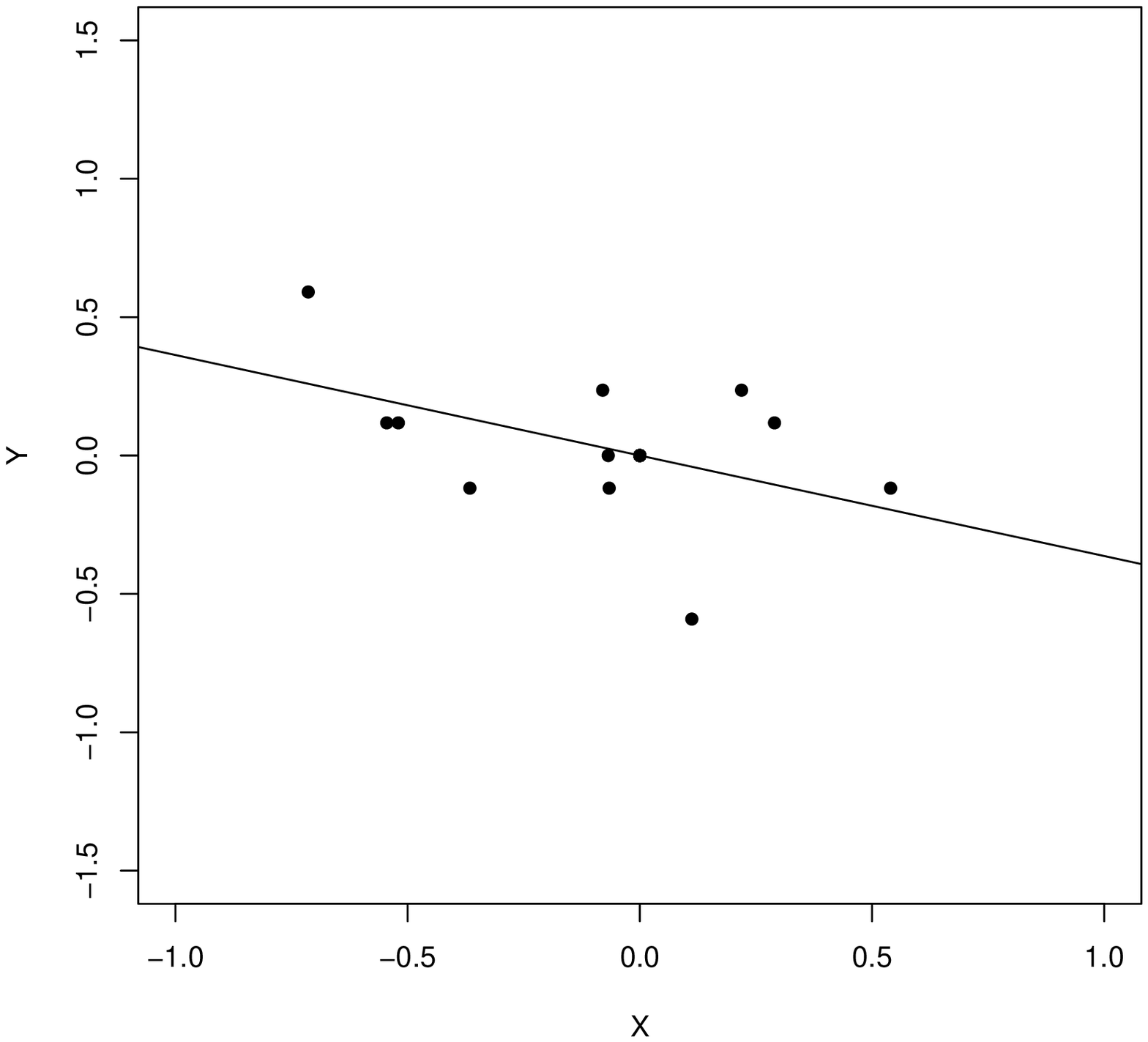}
\caption{The plot whose slope = $-K$ used to determine the Kuramoto coupling parameter. $x$ are the values of $\frac{\partial}{\partial \theta} \big[\rho\int^{2\pi}_{0} \sin(\theta^{'}-\theta)\rho(\theta^{'},t,0) d\theta^{'}]$ while $y$ are the values of $\omega\frac{\partial\rho}{\partial \theta}$}
\label{slopeplot}
\end{figure}

A total of $N=35$ people volunteered for the experiment. A larger sample size was made difficult due to the surprising absence of wrist-watches amongst many solicited volunteers (everyone uses cell phones) and the reticence of others to participate. The distribution of the deviations by minute from the computer clock time are shown in figure \ref{timedistribution}. The error of each measurement is assumed to be $\pm 1$. 
In order to test for the confidence of the mean and variance of the measurements in the sample versus the hypothetical population, the mean and its 95\% confidence interval were calculated using a 1000 sample bootstrap and a two-sided t-statistic with df=34. The bootstrap mean phase was -0.13, a bias of 0.01 versus the sample mean and the standard error was 0.001 so the 95\% 
confidence interval of the mean phase is -0.124 to -0.128. This range is below the phase difference for the one minute error uncertainty of 0.16 and shows that the sample is likely a close representation of the whole population. To estimate the standard deviation a simple jacknife procedure was used. 
The sample standard deviation was 0.25 while the jacknife standard deviation returned 0.24 with a standard deviation of this estimate being 0.006. Given the normality of the original data, these tight estimates give us a reasonably reliable confidence that the phase distribution is close to that of the population
the sample was drawn from.

Next, the order parameter, $r$, was calculated and given the propagation of the measurement error $r$ was determined to be $0.97 \pm 0.02$. Next the mean coupling was derived assuming that the measurements represented a sample of an infinite-N Kuramoto coupled oscillator network with a phase density, $\rho$ where

\begin{equation}
\int^{2\pi}_{0} \rho(\theta,t,\omega)d\theta = 1
\label{rho}
\end{equation}

A continuity equation between $\rho$ and the instantaneous velocity of the oscillators $v$ is given by

\begin{equation}
\frac{\partial \rho}{\partial t} = -\frac{\partial}{\partial \theta}(\rho v)
\label{continuity}
\end{equation}

Strogatz and Mirollo determined in \cite{infiniteN1,infiniteN2} the relationship between the phase density and the oscillator frequency distribution $g(\omega)$ can be given by

\begin{equation}
\frac{\partial \rho}{\partial t} = -\frac{\partial}{\partial \theta} \bigg[ \rho \bigg( \omega + K\int^{2\pi}_{0} \int^{\infty}_{-\infty}\sin(\theta^{'}-\theta)\rho(\theta^{'},t,\omega^{'})g(\omega^{'})d\omega^{'}d\theta^{'}\bigg)\bigg]
\label{bigeq1}
\end{equation}

Since all the oscillators have the same frequency, $g(\omega)$ can be modeled by a Dirac delta distribution $\delta{\omega-\omega_0}$ and the equation changes to

\begin{equation}
\frac{\partial \rho}{\partial t} = -\frac{\partial}{\partial \theta} \bigg[ \rho    \bigg( \omega + K\int^{2\pi}_{0} \int^{\infty}_{-\infty}\sin(\theta^{'}-\theta)\rho(\theta^{'},t,\omega^{'})\delta(\omega^{'}-\omega_0)d\omega^{'}d\theta^{'}\bigg)\bigg]
\label{bigeq2}
\end{equation}

Integrating the delta distribution we simplify to

\begin{equation}
0 = -\frac{\partial}{\partial \theta} \left[\rho \left(\omega + K\int^{2\pi}_{0} \sin(\theta^{'}-\theta)\rho(\theta^{'},t,0)) d\theta^{'} \right) \right]
\label{bigeq3}
\end{equation}

And finally

\begin{equation}
\omega\frac{\partial\rho}{\partial \theta} = -K\frac{\partial}{\partial \theta} \left(\rho\int^{2\pi}_{0} \sin(\theta^{'}-\theta)\rho(\theta^{'},t,0) d\theta^{'} \right)
\label{bigeq4}
\end{equation}

From this final equation we can then use the data collected to determine the phase density and to calculate the derivatives and integrals by differences or summations over multiple 1 minute ($\frac{2\pi}{60}$) phase steps. Finally by plotting the derivative on the left and the expression on the right, we get a curve whose slope is equal to $-K$. In figure \ref{slopeplot} the quantities are plotted using the phase data derived from the experimental observations. The value of $K$ is estimated at 0.36 with an $R^2$ of 0.37 and a 95\% confidence interval of [0.24,0.49]. This estimated $K$ combined with the calculated $r$ compares favorably for example with numerical simulations of $r$ and $K$ on scale-free networks \cite{scalefree} where $r =0.97$ at approximate $K\approx0.4$. In addition for a Dirac delta distribution, the calculation of $K_c$
\begin{equation}
K_c =\frac{2}{\pi g(0)}
\label{Kc}
\end{equation}

gives a result of $K_c=0$ so any amount of coupling would lead to a collective synchronization in the watch network.

\section{Discussion}

The Kuramoto model is the simplest model of oscillator networks with no direct regard to complex topologies. In more complex networks, the network of interactions does matter and can affect the speed of the onset of synchronization stability as well as the critical coupling and behavior near the critical coupling point \cite{netsyncreview}. In order to move beyond using dynamics to derive a value for the coupling under the Kuramoto model, it would be necessary to demonstrate deviations in the obtained data from `ideal' Kuramoto behavior under the constraints defined for the system.  However, this is more easily done by being able to view networks at various states of synchronization and coupling rather than network dynamics already at steady state as shown in this paper. In particular, observing the onset of the critical coupling in a real network and comparing it to its Kuramoto ideal can allow you to extract valuable information such as the index eigenvalue which can provide an bounds on the diameter of the network following spectral graph theory.

Unfortunately, such rich data was not available in this study and thus only a tentative measure of the Kuramoto coupling was established. In the future, the hope is that detailed measures of network synchronization over time, for example in power grids, neurons, or organisms, can allow us to derive a rough idea of the topology of interaction amongst the given components. With this information, deductive reasoning can be employed to link the likely structure with function and interaction mechanisms. Thus the loop will be closed and a key question about the interactions of topology and dynamics will bear fruit.
\ack
The author would like to thank the personnel of the Rochester Public Market, particularly Joan Hildebrand, for their kind help in permitting and setting up the survey table and George Hrabovsky for assistance on the paper.

\end{document}